\documentclass[conference]{IEEEtran}
\usepackage{amssymb}
\usepackage{amsthm}
\usepackage{eucal}
\usepackage{eufrak}
\usepackage{dsfont}
\usepackage{graphics}
\usepackage[pdftex]{graphicx}
\usepackage{multirow}
\usepackage{rotating}
\usepackage{amsmath, amssymb}

\newtheorem{theorem}{Theorem}
\newtheorem{corollary}{Corollary}

\newtheorem{remark}{Remark}

\newtheorem{definition}{Definition}

\newcommand{\comment}[1]{}

\newcommand{\PP}{\mathds{P}}

\newcommand{\toinf}{\rightarrow~+\infty}
\newcommand{\limT}{\lim_{T\rightarrow~+\infty}}

\newcommand{\hl}{\hat{\lambda}}
\newcommand{\tl}{\tilde{\lambda}}
\newcommand{\hr}{\hat{\rho}}

\newcommand{\ha}{\hat{\alpha}}

\newcommand{\haa}{\hat{a}}

\newcommand{\reqA}{\textbf{Requirement-A}}
\newcommand{\reqB}{\textbf{Requirement-B}}
\newcommand{\reqC}{\textbf{Requirement-C}}

\begin{document}
\title{Rate based call gapping with priorities and fairness between traffic classes}

\author{Benedek Kov\'acs$^{\mathrm{a},\mathrm{b}}$,\\
\small\itshape $^a$Ericsson Research Hungary\\ $^b$ Department of
Mathematical Analysis, Budapest University of Technology and
Economics H-1111, Egry J\'ozsef u. 1, Hungary\\
E-mail: benedek.kovacs.mr@gmail.com}

\comment{
\author{Benedek Kov\'acs\\
Ericsson Research Hungary and\\Budapest University of Technology and Economics\\
Email: benedek.kovacs.mr@gmail.com\\
NOT FINAL VERSION}}

\maketitle

\begin{abstract}
This paper presents a new rate based call gapping method. The main
advantage is that it provides maximal throughput, priority handling
and fairness for traffic classes without queues, unlike Token Bucket
which provides only the first two or Weighted Fair Queuing that uses
queues.

The Token Bucket is used for call gapping because it has good
throughput characteristics. For this reason we present a mixture of
the two methods keeping the good properties of both.

A mathematical model has been developed to support our proposal. It
defines the three requirements and proves theorems about if they are
satisfied with the different call gapping mechanisms. Simulation,
numerical results and statistical discussion are also presented to
underpin the findings.
\end{abstract}

\section{Introduction}
There are many overload and load sharing problems to be solved in
telecommunication networks of various kind e.g. in the Internet
Multimedia Subsystem. Considering any type of network and signalling
protocol a protocol operation flow consists of \textit{messages}.
The network nodes are entities receiving these messages and they
process them using their resources such as CPU capacity or memory.
In case they lack the resource to process the message we say that
the node is \textit{overloaded}.

\begin{figure}[h]
\begin{center}
\includegraphics[width=0.45\textwidth]{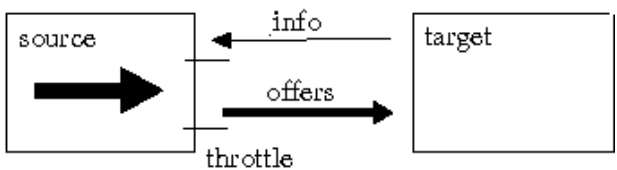}
\end{center}
\caption{Schematic architecture of traditional external overload
control mechanisms. The throttle entity is part of the source
node.}\label{fig:ovl-ctrl-architecture}
\end{figure}

To avoid such situations the node itself can deny to serve the
request and reject (send a negative reply message) or drop (ignore)
it. Another solution can be that the sender (\textit{source}) does
not send out the message if for some reason it knows that the
\textit{target} will not be able to serve it. In both cases there is
a decision logic deciding upon admission of the request or sending
out the request (see Figure~\ref{fig:ovl-ctrl-architecture}). This
entity is called the \textit{throttle} which is in the center of our
interest. From now on we use the following terminology and model
(see Figure~\ref{fig:throttle-entity}).

\comment{
\begin{figure}
\begin{center}
$\begin{array}{lr}
\includegraphics[width=0.45\textwidth]{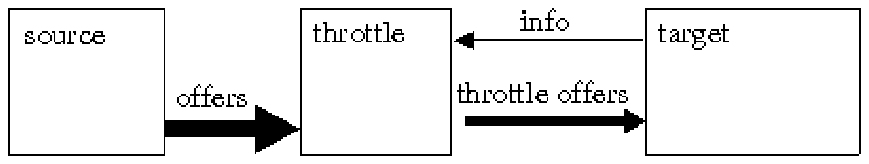}
&
\includegraphics[width=0.45\textwidth]{terminology-concept.eps}
\end{array}$
\end{center}
\caption{The throttle as an individual
entity.}\label{fig:throttle-entity}
\end{figure}
}

\begin{figure}[h]
\begin{center}
\includegraphics[width=0.45\textwidth]{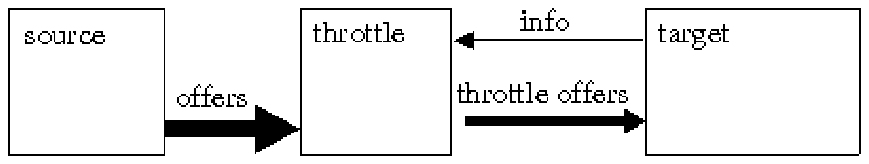}
\end{center}
\caption{Schematic architecture of our model. The throttle is an
individual functional entity. It can classify incoming offers and
decides on admission or rejection of the
offer.}\label{fig:throttle-entity}
\end{figure}

Definitions of the system elements and technical assumptions for the
model:
\begin{itemize}
 \item The \textit{throttle} decision function is a function mapping from the offer load point process
 to the set \{\textit{admission}, \textit{rejection}\}. (Each
 \textit{throttle} is uniquely assigned a function $\gamma$ that transforms the intensity process $\rho(t)$ of the income process to
an intensity process of the admissions
$\gamma(\rho(t))=a(t)$.~\cite{OGATA})
 \item An \textit{offer} is the event for which the \textit{throttle} has to decide on \textit{admission} or \textit{rejection}. If an offer is
admitted it cannot be rejected (dropped) and vice versa, and there
is no third possibility. An offer has properties: \textit{arrival
time, priority level and class} which can be measured.
 \item The \textit{traffic class} and the \textit{priority level} sets have finite elements.
 \item The \textit{offered traffic} (or \textit{offer load}) is the flow of \textit{offers} modeled with a
progressively measurable not necessarily stationary point process
marked with the marks from the mark space that is the direct product
of the set of priorities and classes. (This implies that the
probability of two \textit{offer} events occurring at the same time
is zero.)
 \item The \textit{admitted traffic} (or \textit{throughput}) is the flow (i.e. the point process) of \textit{admitted offers}
(\textit{offers} for which the \textit{throttle} yields admission).
The flow of admitted offers can be conditioned upon the whole
history (past) of the offer load flow and upon the throttle
parameters and of course on the decision strategy.
\end{itemize}

The above assumptions and definitions are natural and obvious and
also necessary to make the discussion clear.

The \textit{throttle} entity discussed here is one very important
and well defined part of overload control systems of any type as it
has the role to reject (or drop) an \textit{offer} or to let it go
through: admit it. The \textit{throttle} realizes a call gapping
mechanism if it makes the decision based only on previous offers
i.e. no offers in the future are examined. This also means that in
our case the non-anticipative \textit{throttle} is not allowed to
delay an \textit{offer} and only one \textit{offer} arrives at a
time i.e. the call gapping mechanism cannot buffer the offer and
admit it later than it has arrived. This makes a fundamental
difference from Weighted Fair Queuing and mechanisms like those
in~\cite{WFQ,GPS}.

Many call gapping mechanisms have been developed for different
purposes with different characteristics. One of the most important
call gapping algorithms is the Crawford algorithm~\cite{CRAWFORD}.
It does not differentiate between incoming offers. One of the most
common solutions that handles priority levels and some kind of
traffic classes is the Token Bucket call gapping
mechanism~\cite{H.248.11}. This one is also popular because it is
also used to characterize telecommunication traffic~\cite{TBCHAR}.

\comment{(This work is motivated by the problem of providing
resource sharing in telecommunication networks with call gapping in
case of overload. The problem arises when concurring applications
use the same limited resource in the system i.e. processor capacity,
memory. The problem to be solved is similar to the ones solved by
Fair Queuing, Weighted Fair Queuing, General Processor
Sharing~\cite{WFQ,GPS} but the requirements and the environment is
different. The fundamental difference is that there are no queues
allowed in our case and it is desired to keep some kind of Token
Bucket property.)}

The aim here is to present a traffic estimation based call gapping
mechanism that can provide traffic share Service Level Agreement,
like weighted fair queuing mechanisms but without queuing the
traffic. We discuss the following requirements for a call gapping
rate limiting \textit{throttle} mechanism. The typical verbal
definitions given here preliminary, are not precise and many
contradict and can have multiple exact (i.e. mathematical)
definitions with different results\comment{(they are polysemantic)}.

\begin{itemize}
 \item \reqA\label{req-max-throughput}\ \textit{Maximal
throughput with bound}: No \textit{offer} should be rejected if
there is enough available capacity in the system to serve it, but no
\textit{offer} should be admitted if there  is not enough available
capacity to serve it in the system.
 \item \reqB\label{req-priority-calss}\ \textit{Priority
levels}: Each \textit{offer} may be assigned a priority level and
the \textit{offer} with higher priority shall be admitted in favor
of the one with the lower priority level.
 \item \reqC\label{req-throughput-share}\
\textit{Throughput share for traffic classes}: The offers can be
classified and for the traffic class $i$ the $s_i$ portion of the
capacity of the target shall be provided.
\end{itemize}
In this paper we give exact definitions of these requirements.

In Section~\ref{section:token-bucket-throttle} we show how the Token
Bucket mechanism meets \reqA\ and then for \reqB. Token Bucket does
not fulfill \reqC. In
Section~\ref{section:new-call-gapping-mechanism} our new method is
presented. We give mathematical definitions of all the requirements
and prove that our new method meets \reqA\ and \reqC. Then we
present a call gapping method that is a mixture of the latter two
and also show that it meets the requirements. In
Section~\ref{section:numerical-results} we present our simulations
and some figures about the offer and admission traffic flows with
the three mechanisms. Using statistics we show how each mechanism
meets \reqB.

\section{Token Bucket Throttle}\label{section:token-bucket-throttle}
We do not want to go into details discussing the throughput
regulation properties of a Token Bucket algorithm (defined e.g. in
patent~\cite{TBLBW} and used e.g. in standard~\cite{H.248.11}), but
it is necessary to give a brief description to underpin the
assumptions of our model. At first we present the concept of Token
Bucket then show how it was extended to meet \reqB.

\subsection{The Token Bucket with parameters $(r,W)$}
The Token Bucket call gapping mechanism is the following: there is a
bucket of available tokens representing available resources (free
capacity) of the system. Requests are offered to the system and each
of them is assigned a number of tokens needed i.e. the amount of
resources it requires to be served. Once there are enough tokens in
the bucket the request is admitted and dropped otherwise. (Thus no
queues are applied and no delay is present in the system because of
the Token Bucket call gapping algorithm.)

By the definition of the original Token Bucket the tokens are
generated into the bucket with exponential distribution and the
offers arrive with a Poisson process in most models that means that
the time interval between the arrivals is also exponentially
distributed. We analyze and describe a variant of this.

At first we mention that decision about serving a request are often
implemented differently. The most important difference is in the
interpretation: rather than consuming the tokens the bucket fill $b$
is increased when a request arrives. The token generation is then
realized with decreasing the bucket fill. The maximum fill is the
watermark $W$ that cannot be exceeded and also the bucket fill can
not be lower than 0. This concept is equivalent to the original
algorithm.

Secondly we consider deterministic token generation instead of the
exponential one that is used in most cases (e.g.~\cite{H.248.11}),
because it is much easier to implement and sometimes to analyze, as
well.

Then the Token Bucket mechanism we discuss works as follows: When a
new request arrives at $t_n$ than the needed bucket fill is
calculated: $b(t_n)$ as if the request was served. This is done with
calculating the expected number of tokens that would have been
generated from the time the former service was served ($t_{n-1}$)
then multiply it with the throughput capacity of the bucket i.e. the
Token Bucket rate at $t_n$: $r(t_n)$ and subtract it from the former
bucket size at $b(t_{n-1})$. Then it tests it against the preset
constant watermark: $W$.
\begin{definition}[Token Bucket call gapping strategy
$\gamma_t(r,W)$]
\begin{equation}\label{eq-bucketfill-basic}
b(t_n)=\max\{\chi(t), b(t_{n-1})-r(t_{n-1})(t_n-t_{n-1})+\chi(t)\},
\end{equation}
where $\chi(t)=1$ iff there is an offer. Admit if $b(t_n)\leq W$. If
the offer is admitted, the above definition is used for the next
value of the bucket fill $b$. If the offer is rejected, then
$b(t_n)$ is recalculated with $\chi(t)=0$.
\end{definition}

(In many solutions the offers for the bucket can be of different
types with different resource needs and thus $S_i \chi_i(t)$ is used
for update, where $S_i$ is the so-called ``splash amount'' i.e. the
expected number of tokens needed to serve the request of type $i$.
From now on we suppose that $S_i=1$, since the calculations would be
much more difficult without any qualitatively different result with
respect to the requirements we consider now.)

\subsection{Priority handling with Token Bucket}
Once the offered traffic is modeled with a point process and the
throttle meets \reqA\ we cannot provide priority between the offers.
Why? Suppose that we have an offer in the system and we have to
decide if we should admit it or not. \reqA\ tells us to admit the
offer if we have the capacity to serve it. Suppose that this is the
case and see that if the throttle would not admit the current offer
to reserve this capacity for offers of higher priority then it might
happen that there will be no higher priority offer in the future and
the throttle would suffer a loss of workload.

\comment{It is not so obvious problem if the capacity is described
with two parameters that cannot be compared i.e. Token Bucket with
$r,W$.}

However, giving up the maximal throughput requirement some priority
handling naturally can be done. In the Token Bucket concept
different watermarks are assigned to each priority level. The offers
of lower priority are checked with a lower watermark. This is kind
of reserving a set of tokens (system resources) to the higher
priority traffic. This method violates \reqA\ whenever $b(t)$
declines to $0$ before rejecting an offer. Whenever this event has a
low probability, using different watermarks for different priority
levels is a good solution to meet \reqB\ with a Token Bucket
throttle.

\section{Call gapping with rate estimation}\label{section:new-call-gapping-mechanism} In this
Section our new method, the proposed rate based call gapping
throttle is presented. At first we introduce the complete proposed
procedure clearly. Then we discuss and prove how it provides all the
requirements and what possible extensions, modifications or other
solutions might result a similar good algorithm. At the end of the
discussion we present relationship between the new method and the
original Token Bucket algorithm.

\subsection{The new call gapping algorithm $\gamma_g(c,T,g,s)$}
Suppose that the consecutive offers arrive to the \textit{throttle}
at $...<t_{n-1}<t_n<t_{n+1}<...$ time instants respectively. Each
\textit{offer} has a well defined \textit{priority level}
$j,j\in1..J$ and \textit{traffic class} $i\in1..I$. Each
\textit{priority level} $j$ has a constant priority parameter $T_j$
assigned ($T_j\geq T_k)$ if the \textit{offer} with priority $k$ has
the higher priority) and each \textit{traffic class} has a
pre-configured weight $i\mapsto s_i\in(0,1)\subset\mathds{R},$
(where $\sum s_i=1$). For each $i$ the algorithm maintains an
estimation of the incoming \textit{offer rate} $\hat{\rho}_i(t)$, a
\textit{provisional admission} rate $\hat{\alpha}_i(t)$ from which
it calculates a \textit{bounding rate} $g_i(t)$ and then according
to the decision it estimates an \textit{admission rate}
$\hat{a}_i(t)$.

We suppose that the rate of the throttle varies with the following
function: $c(t)$. (This value is determined and given for the
algorithm and represents the capacity of the \textit{throttle} and
might be different from $r(t)$). \comment{We propose that the timer
$T_j:=K/c(t)$ where $K$ is an arbitrary constant that we will
explain later.}

\begin{definition}[The rate based call gapping $\gamma_g(c,T,g,s)$.] Define the proposed throttle decision strategy
$\gamma_g$ in the following way. Suppose that at $t_n$ an
\textit{offer} arrives and the system is in state
$\{t_{n-1},\hat{\rho}_i(t_{n-1}),\hat{a}_i(t_{n-1})\}$ and $c(t_n)$:
\begin{enumerate}
  \item Determine priority constants, i.e. calculate $T_j$;
  \item Update the incoming rates estimate for all $i$:
  $\hr(t_n)$ with $\chi_k(t_n)=1$ iff $i=k$, 0 otherwise;
  \item Calculate a provisional admission rate for all $i$:
  $\ha(t_n)$ with $\chi_k(t_n)=1$ iff $i=k$, 0 otherwise;
  \item Calculate the bounding rate for class $i$ only: $g_i(t_n)$;
  \item If $\hat{\alpha}_i\leq g_i$ then \textit{admit} the offer and
  $a(t_n):=\alpha(t_n)$ else \textit{reject} the
  \textit{offer} and update $\haa(t_n)$ with $\chi_k(t)=0,\forall k$(!);
  \item (Continue with 1. for the next event).
\end{enumerate}

We propose to update $\hat{\rho}_i,\hat{\alpha}_i,\hat{a}_i$
according to the following equation:
\begin{equation}\label{eq:rate-estimator}
\hat{\lambda}(t_n):=\frac{\chi(t_n)}{T_j}+\max\{0,\frac{T_j\hat{\lambda}(t_{n-1})-(t_n-t_{n-1})\hat{\lambda}(t_{n-1})}{T_j}\},
\end{equation}
where $\hat{\lambda}$ is an estimator asymptotically unbiased for
the $\lambda(t)$ real intensity of a point process thus to be
replaced by $\hat{\rho}_i,\hat{\alpha}_i,\hat{a}_i$ and indicator
$\chi_i(t_{n-1})=1$ iff the \textit{offer} is of type $i$ and 0
otherwise (or further specified like in step 5). Note that the time
parameter $T_j$ changes in time too according to the priority level
and the former one always has to be remembered.

To calculate the bound rate at first we introduce $u(t)$ the
provisional used capacity according to \reqB:
\begin{eqnarray}\label{eq:unused-capacity}
u(t)&:=&\sum_{\forall i}\min\{s_i
c(t),\hat{\rho}_i(t)\}\\
&=&\sum_{\hat{\rho}_i(t)\leq s_i c(t)}\hat{\rho}_i(t)+\sum_{s_i
c(t)<\hat{\rho}_i(t)}s_i c(t)\nonumber
\end{eqnarray}
Then the remaining (unused) capacity in the system is $c(t)-u(t)$.
This has to be split between traffic classes with higher incoming
rate then the agreed share $\hat{\rho}_i(t)>s_ic(t)$. Then
\begin{equation}\label{eq-g-new-proposal}
g_i(t):=\min\{\hat{\rho}_i(t),s_i c(t)+(\hat{\rho}_i(t)-s_i
c(t))\frac{c(t)-u(t)}{\rho-u(t)}\}.
\end{equation}
\end{definition}

It is important to see that our method is capable to handle other
class-wise throughput criteria than fair sharing and maximal
throughput. Giving upper or lower bounds for $g$ one can implement
fairly complex throttle mechanisms.

As one can see the new method is more complex than the original
token bucket mechanism. However, the processing cost of updating the
few variables introduced is significantly smaller than processing
the offers thus does not count even in case of overload.

\subsection{$\gamma_g$ meets all the
requirements}\label{section:g-meets-the-requirements} Now that the
strategy is introduced we prove that it meets all the requirements.
At first we define each requirement mathematically then we show how
they are satisfied. We introduce some notation to make the
discussion clear.
\begin{itemize}
 \item $c(t)$ represents the true capacity of the system expressed in rate, i.e. some
 deterministic value coming from an external input source.
 \item $\rho(t)$ is the real intensity of the offered traffic and $\hat{\rho}(t)$ its
 estimate with~\eqref{eq:rate-estimator}.
 \item $a(t)$ is the real intensity of the admitted traffic and
 $\hat{a}(t)$ is the estimation of the rateintensity with~\eqref{eq:rate-estimator}.
 \item $\alpha(t_n)$ is the preliminary admitted traffic intensity for which
 the following stands: $\alpha(t)=a(t),\forall t<t_n$ and
 $\alpha(t_n)$ is the intensity $a(t_n)$ would have if the offer was
 admitted at time $t_n$, and its estimate is $\hat{\alpha}(t)$
 accordingly.
\end{itemize}

\subsubsection{Requirement-A\ }
This requirement consists of two parts. At first it says that there
exists an upper bound for the system that should not be exceeded,
i.e. it limits the admission rate to avoid overload. Secondly, it
tells us that once the limit is not exceeded then all the offers
should be admitted to maximize the utilization. However, in theory
the words capacity and bound can have many different definitions
depending on the model we use for the target node.

The target node is often modeled with an inverse Token Bucket, i.e.
server with deterministic serving rate $s$ and a queue of maximal
length $Q$. It is very easy to see that the Token Bucket throttle
$\gamma_t(s,Q)$ can perfectly meet the requirement in this case.
(Note that this is true supposed that there is no delay in the
system between the throttle and the protected entity while
$s(t)=r(t)$ is satisfied.)

Another approach is to assume that the target can handle requests on
a maximal call rate $c$ that is used as the bound at the throttle.

Both models have benefits and drawbacks while a mixture of them is
used in practice. Speaking about the capacity of a node in Next
Generation Networks engineers often refer the call rate value in
industrial contracts and Service Level Agreements. It is very
important to note that the feedback driven overload control
mechanisms work with call rate information too (see \cite{H.248.11}.
On the other hand a server with queue is a common model in the
academic literature for the CPU capacity and Token Bucket (or
versions of it) is proposed in many standards (e.g. \cite{H.248.11}
again) and implemented into nodes.

As a consequence we say that although it is rather difficult to give
exact definition for \reqA\ we can give some definition grabbing a
few properties depending on the method we use.

\begin{definition} Call rate bound. \reqA\ is met if $\sum
E[a_i(t_n)]\leq c(t_n)$ (the throughput rate is bounded in expected
value).
\end{definition}

\begin{theorem}\label{thm-gammaG-maxbound}
The throttle with strategy $\gamma_g$ meets the \textbf{call rate
bound} requirement.
\end{theorem}

\begin{proof} The proof relies on the fact that the estimator is asymptotically unbiased i.e. $\limT
E[\hat{a};T_i]=E[a_i]$ with negative bias if $T\geq1/a_i$ (thus
$E[\hat{a}_i]<E[a_i]$). The proposed strategy $\gamma_g$ limits
$a_i$ so that $a_i\leq g_i$ thus we are ready if we show that
$g(t):=\sum g_i(t)=c(t)$.

Define $u_1(t):=\sum_{i:\hat{\rho}_i(t)<s_i c(t)}\hat{\rho}_i(t)$
and $u_2(t):=\sum_{i:s_i c(t)<\hat{\rho}_i(t)}s_i c(t)$ thus
$u=u_1+u_2$ and then $g_i=\min\{\hat{\rho}_i,s_i
c+(\hat{\rho}_i(t)-s_i c(t))\frac{c-u}{\rho-u}\}$. Although the
system is non-stationary it is homogenous in time so
$f(t)=\mathrm{const.}$ for all functions. Now calculate $g(t)$):

\begin{eqnarray}\label{eq-extended-fairness-proof}
g&=&\sum
g_i=\sum\min\{\hat{\rho}_i,s_ic+(\hat{\rho}_i-s_ic)\frac{c-u}{\rho-u}\}=\nonumber \\
&=&\sum_{i:\hat{\rho}_i<s_ic}\hat{\rho}_i+\sum_{i:s_i
c<\hat{\rho}_i}s_i
c+(\hat{\rho}_i-s_ic)\frac{c-u}{\rho-u}=\nonumber \\
g&=&u_1+u_2+(\rho-u_1-u_2)\frac{c-u_1-u_2}{\rho-u_1-u_2}=c.
\end{eqnarray}
\end{proof}

\begin{corollary} The following calculation of $g$ can also be used:
\begin{equation}
g_i(t)=\min\{\hat{\rho}_i,s_i c(t)+(\hat{\rho}_i(t)-s_i
c(t))\frac{c(t)-u(t)}{\rho-u(t)}\},
\end{equation}
where $u(t)=\sum_{i:\hat{\rho}_i(t)<s_i c(t)}\hat{\rho}_i(t)=\alpha(t)$.
Then~\eqref{eq-extended-fairness-proof} becomes:
\begin{equation}
g'=u_1+u_2+(\rho-u_1-u_2)\frac{c-u_1-u_2}{\rho-u_1-u_2}=c.
\end{equation}
\end{corollary}

The difference between the two strategies is that in case of $g$ the
remaining capacity is split between the classes with higher offer
rates proportionally to their weights while using $g'$ the remaining
capacity is split proportionally the remaining offer rates. Both
satisfies \reqA\ and as we will see \reqC. From now on $g$ means
either $g$ or $g'$ and the results will be the same obviously.

\subsubsection{Requirement-B}
As pointed out before, the priority requirement for call gapping is
the most complex in a way since in the gapping algorithms it is
supposed that we make decisions using measures on the past and the
present offer. No future events can be used thus \reqB\ is always
satisfied. There is always one offer in the system and the throttle
can admit or reject it according to \reqA\ and \reqB.

In case of the Token Bucket call gapping different watermarks $W_j$
are introduced for each priority level $j$. One interpretation is
that the bucket allows larger peaks for traffics with higher
priority thus $W_j<W_k$ whenever $k$ represents the higher priority
level. Doing this, the bucket implicitly reduces the throughput for
lower priority traffics (the extra peak in the bucket has to be
refilled with tokens i.e. $b(t)$ has to decline below the low
watermarks to admit low priority traffic). Note that the different
watermark levels has no effect if the offer rate is low with small
peaks thus the rejection probability is small i.e. if there is no
overload. Supposed that the true bound is $W=\max\{W_j\}$ this
system preserves capacity for high priority traffic.

We give a similar solution for the problem through the timer
parameter of the estimators: $T$. As it was defined we introduce a
function of $T:j\mapsto T_j$ where $T_k\leq T_j$ if $k$ represents
the higher priority. (Note that it is the other way around for
$W_j$s.) The interpretation is that the estimator forgets the high
offer rates faster for the traffic of the higher priority. Let
$T_m=\min\{T_j\}$, the true bound on the throttle using different
$T_j$s, means that for low priority traffic it remembers the high
peaks for a longer period thus reserves capacity for the higher
priorities similarly to the Token Bucket.

The two methods have different characteristics, but one thing is
common. Both reserve capacity for higher priority traffic. Now we
say that to meet \reqB\ the system has to have this ability and
define it in the following way.

\begin{definition}[Requirement on priorities.]
Suppose that the throttle has rejected an offer at time $t_{n-1}$.
Let $t_{n;j}$ be the closest time the throttle is able to admit an
offer of priority level $j$. \reqB\ is met iff $\forall k,l
(t_{n;k}\leq t_{n;l}) \Leftrightarrow (k\geq l)$ ($k$ represents a
higher priority).
\end{definition}

The exact proof of this statement is not ready yet. Simulation
result shows that the proposed strategy satisfies \reqB. We discuss
the statement in the Numerical Results Section.

\subsubsection{Requirement-C}
This is referred to as the throughput share requirement and tells us
that there should be at least an $s_i$ portion of the capacity
dedicated to traffic class $i$.

\begin{definition}[\reqC.] The \textbf{Minimum share
requirement} is met if $\forall i:(\rho_i(t_n)\leq
s_ic(t))\Rightarrow E[a_i(t_n)]=\rho_i(t_n)$ i.e. if the offer rate
of a traffic class is less than the agreed share it should be fully
admitted.
\end{definition}

\begin{theorem}
The throttle with strategy $\gamma_g$ meets \reqC\ in expected
value.
\end{theorem}

\begin{proof} At first we have the asymptotical unbiasedness for our estimators thus
$\limT E[\hat{a;T}_i]=E[a_i]$ thus the proof is true for the
expected value of $a_i$.

Statement $\hat{a}_i(t_n)=\hat{\rho}_i(t_n)$ whenever $\forall i
\hat{\rho}_i(t_n)\leq s_ic(t)$ is equivalent to the statement
($g_i(t_n)\geq \hat{\alpha}_i(t_n)$ thus) $g_i(t_n)\geq
\hat{a}_i(t_n)$ whenever $\hat{\rho}_i(t_n)\leq s_ic(t)$. According
to strategy $\gamma_g$: $g_i(t_n)=\hat{\rho}_i(t_n)$ whenever
$\hat{\rho}_i(t_n)\leq s_ic(t)$ and since
$\hat{\alpha}_i(t_n)\leq\hat{\rho}_i(t_n)$ because
$\hat{a}_i(t_{n-1})\leq\hat{\rho}_i(t_{n-1})$, it is true that
$\hat{\alpha}_i(t_n)\leq g_i(t_n)$ thus the offer is admitted (and
also $\hat{a}_i(t_n)\leq g_i(t_n)$).\end{proof}

\subsection{Rate model for Token Bucket and a joint algorithm merging the
methods}\label{section:token-bucket-rate-model-class-handling} In
this section we introduce a model for Token Bucket that is
equivalent to the definition in
Section~\ref{section:token-bucket-throttle} but makes calculations
easier.

\begin{definition}
Token Bucket Rate Model Strategy: $\tilde{\gamma}_t(r,W)$ Let us
define $T(t)=W/r(t)$ and use the following equation for updating the
bucket rate variable:
\begin{equation}\nonumber
\tilde{a}(t_n)=\frac{\chi(t_n)}{T}+\max\{0,\frac{T
\tilde{a}(t_{n-1})-(t_n-t_{n-1})r(t_n)+}{T}\}
\end{equation} where $\chi(t)=1$ iff there is an offer at time $t$. Admit the offer iff
$\tilde{a}(t_n)\leq r(t_n)$. If the offer is admitted then the above
definition is the used for the next value of the bucket rate
variable $\tilde{a}(t)$. If the offer is rejected then
$\tilde{a}(t_n)$ is recalculated with $\chi(t)=0$.
\end{definition}

\begin{theorem}
The Token Bucket and the Token Bucket Rate Model Strategy are the
same: $\gamma_t=\tilde{\gamma}_t$.
\end{theorem}

\begin{proof} It is easy to show that $b(t_{n-1})=\tilde{a}(t_{n-1}) T
\Rightarrow b(t_n)=\tilde{a}(t_n) T$ and the decision is $b=T
\tilde{a}(t)\leq T r(t)=W$ also trivial.\end{proof}

If one extends the Token Bucket for traffic class handling with some
role like in the proposed mechanism it will not provide traffic
class fairness. The reason is hidden in the fact that unlike
$\tilde{\rho},\tilde{\alpha},\tilde{a}$, $\tilde{\beta}$ and all
such estimators is not asymptotically unbiased i.e.
$E[\tilde{\lambda}]=\lambda$ as $t\toinf$ is not true for the
estimators defined with:
\begin{equation}\label{eq:rate-statistics-bucket}
\tilde{\lambda}(t_n)=\frac{\chi(t_n)}{T_j}+\max\{0,\frac{T
\tilde{\lambda}(t_{n-1})-(t_n-t_{n-1})r(t_n)}{T}\}.
\end{equation}

The bucket fill does not represent at all the used capacity in the
system it only measures the peakedness of the traffic but these
peaks can happen on low offer rates too.

On the other hand, the proposed method does not allow such big
transient peaks in the traffic. Now we aim to make the proposed new
call gapping to behave like Token Bucket. We define the following
strategy that is a mixed architecture.

\begin{definition}
Rate Based Call Gapping with Bucket-type Aggregate Characteristics:
$\gamma_x$ Take all the definition from the new call gapping
mechanism $\gamma_g$ for $\hat{\rho},\hat{\alpha},\hat{a},u,g_i$ and
define $T_j(t)=W_j/r(t)$. Take $W_j$ and the bucket fill change
definition $b$ from the original token bucket $\gamma_t$. Perform
all the steps like in $\gamma_g$ but decide using the following
constraint equation: $\frac{b(t_n)}{W_j}\hat{a}_i(t_n)\leq
g_i(t_n)$.
\end{definition}

We will show numerically that the mixed algorithm behaves like Token
Bucket on aggregate level and meets all the requirements. The source
of the idea comes from the fact that $\hat{a}(t)$ places a strict
bound on the rate thus $\hat{a}(t)\leq r(t)$ is always true as
required. However we decrease the value of $\hat{a}$ and thus allow
peaks in the traffic like Token Bucket does. (See that Token Bucket
$\gamma_t$ allows temporary bounding violation rate-wise unlike
$\gamma_g$ but like $\gamma_x$. The bucket size related to the whole
bucket is a kind of measure of this violation.)

\subsubsection{$\gamma_{t'}$ and $\gamma_x$ and Requirement-A\ }
Here we discuss how the different algorithms meet the maximal
throughput requirement. It is obvious that Token Bucket cannot meet
\reqA\ in the way it was defined before since that definition
assumed that the target has an infinite queue.

We do not aim to give an exact definition to \reqA\ but we derive
relations between the bucket and the estimator based throughput
characteristics. The number of admitted offers i.e. the probability
of admission is in the center of our interest.

The probability of admission for token bucket depends on the offer
rate with the following formula: $1-\mathrm{Erlang}[\rho,r]$. Thus
the probability of losing calls is only defined at given values of
$\rho$.

For rate based call gapping, since the estimator always
overestimates the rate ($\lambda<\hat{\lambda}$) and cuts the
traffic strictly with $c$ the admission rate is always below the
target. But for the same reason it is possible that the offer is
rejected although it could have been accepted according to the
bound. The probability of this is the probability of estimating
higher rate than $c$ while the true offer rate is lower:
$P[\hat{\alpha}>c|\alpha<c]=1-\frac{P[\alpha<c-B[T]]}{P[\alpha<c]}$,
where $B[T]=1/(T(1-F[T])+E[\Delta t|t<T])-\alpha$ is the bias.
(Knowing the exact bias if constant intensity is supposed for the
offer rate, the bound can be modified to have maximal throughput and
strict bound at the same time.)

The two methods can only be compared at a given value of the
intensity. For all those values when the value of the intensity is
not between $c-B[T]$ and $c$ the $\gamma_g$ strategy works
perfectly. The Token Bucket drops a call with positive probability
for any value of the offer rate and also might admit when the
intensity is higher than allowed. This means that we cannot tell
which method is better or has the higher throughput since it depends
very much on the offer rate.

\comment{Annak lenne itten ertelme, hogy mondjuk $\lambda\in
\mbox{Normal}(c,\mu)$ es igy ki lehet szamolni, hogy hogyan talalja
el a dolgokat egyik, ill. masik. \comment{Arrol van itt szo, hogy a
valaki megmondana nekem, hogy pontosan hova esik az offer rate,
milyen tartomanyba. Mondjuk valaki adna nekem egy eloszlast az offer
rate-re, akkor mar meg tudnam mondani, addig nem.}}

\begin{theorem}\label{thm-gammaX-maxbound} The mixed strategy $\gamma_x$ meets
\reqA\ with appropriate watermark settings.
\end{theorem}

\begin{proof} It is shown in Theorem~\ref{thm-gammaG-maxbound} that
$\sum g_i(t)=g(t)=c(t)$ and since the definition of $g$ was not
changed we should only examine what means to compare $g_i$ to
$\frac{b}{W_j}\hat{a}$ rather than to $\hat{a}$.

When we admit a request then $1\leq b(t_n)\leq W_j\leq W_{\max}$
thus $\frac{1}{W_{\max}} \hat{a}_i\leq\frac{1}{W_j}
\hat{a}_i\leq\frac{b(t_n)}{W_j}\hat{a}_i\leq\hat{a}_i$. This tells
us that $\gamma_x$ lets through more messages than $\gamma_g$ since
$E[\frac{b(t_n)}{W_j}\hat{a}_i]\leq E[\hat{a}_i]$. Fortunately the
maximal watermark limits this overflow error $\frac{1}{W_{\max}}
E[\hat{a}_i]\leq E[\frac{b(t_n)}{W_j}\hat{a}_i]$. It tells us that
there is a setting of watermarks that guarantees bounding. (It is
obvious that if $W_{\max}\toinf$ then $\frac{b}{W_{\max}}\ha$
becomes very small and we always admit the request thus the theorem
cannot be proved for any watermark settings.) \end{proof}

\comment{ AN IDEA: Show that $r<\lambda\Rightarrow\PP[b<W]=0$.
Obviously for $r<\lambda\Rightarrow\PP[b=W]=0$ so this algorithm has
better throughput than the normal rate based call gapping.

This is the reason why simulations shows that this algorithm has the
best throughput.}

\subsubsection{$\gamma_x$ and Requirement-B}
Some simple theorems are proved to show that the mixed strategy
meets the priority and the throughput share requirements.

\begin{theorem}
Token Bucket strategy $\gamma_t$ meets \reqB.
\end{theorem}

\begin{proof} Obviously, the time to accept the next offer of priority level
$j$ is the time when the bucket level declines sufficiently to
$b(t)\leq W_j$. For all levels $k>j$, $W_k>W_j$ i.e. $b(t)$ declines
under the lower threshold later in time and the requirement is
met.\end{proof}

Again it is rather hard to show that the mixed strategy $\gamma_x$
meets \reqB. However, it seems to be trivial that $\gamma_x$
satisfies \reqB\ more drastically than $\gamma_t$ does. We have
interesting simulation results presented about this property. We can
see numerical results about this in
Section~\ref{section:numerical-results}.

\subsubsection{$\gamma_x$ and Requirement-C}
\begin{theorem}
The mixed strategy $\gamma_x$ meets \reqC.
\end{theorem}

\begin{proof} As pointed out $\gamma_x$ admits at least all the offers
$\gamma_g$ does since $\forall i,
\frac{b}{W}\hat{\alpha}_i\leq\hat{\alpha}_i$ is compared to $s_i c$
while a comparison of $\hat{\alpha}$ would be enough. This means
that the mixed strategy provides minimum throughput share and
fulfills \reqC.
\end{proof}

\section{Numerical results and
analysis}\label{section:numerical-results}
Although we have nice proofs on the good behavior of the proposed
rate based call gapping mechanism the complete mathematical
discussion about the differences and similarities with Token Bucket
is not ready yet. It is also true that the requirements can be
interpreted with definitions slightly different from those we gave.
Therefore we would like to present some simulation results and show
that the findings are valid.

The simulation is written in \textit{Mathematica}~\cite{MATHEMATICA}
and a \textit{notebook} is available at
\textit{http://www.math.bme.hu/~kovacsbe/rbcg/BENEDEK-KOVACS-rate-based-call-gapping-PRELIMINARY-VERSION.nb}
as an electronic appendix. \comment{One can also download
demonstrations from the internet page of \textit{Wolfram Inc.}.}

\subsection{\reqA}
The figures shows that all the mechanisms limit the admitted offer
rate while try to keep the highest throughput. In this scenario we
examine the traffic on aggregate level i.e. there is only one
traffic class for which the capacity of the throttle should be
maximized and limited. The capacity is 1 offer/sec for the simple
simulation case while the average number of offers per sec raises
from 0.8 to 2 meaning that there is a 200\% load on the node.

\begin{figure}[h]
\begin{center}
\includegraphics[width=0.45\textwidth]{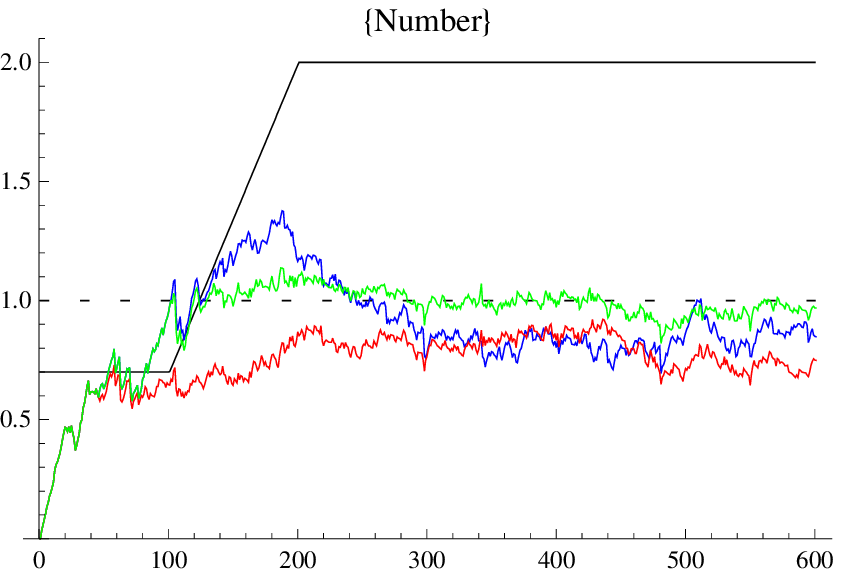}
\end{center}
\caption{The new algorithm ($\gamma_g$) on aggregate
level}\label{fig:throughput}
\end{figure}

As it can be seen in Figure~\ref{fig:throughput} all three
mechanisms limit the admitted traffic although Token Bucket allows
considerable peak at the beginning. (The size of the peak depends on
the parameters we set. Here the 1 offer/sec capacity is very small
compared to the watermark what is set to 10.) On the other hand,
rate based call gapping seems to under-utilize the system while the
joint mechanism seems to have the smoothest and also maximal
throughput.

After a total 600 offers from each traffic with the same exact
trajectory the results shows that $\gamma_t,\gamma_g,\gamma_x$ has
admitted 415, 386, 404 number of calls respectively.

The problem with the mathematical discussion of maximal throughput
is that the results depend very much on the value of the offer rate
and capacity. It is only possible to compare the mechanisms at given
rates what is not available in the world.

\subsection{\reqB\ }
To discuss \reqB\ we provide the reader with some statistical
results. The sample is generated with our simulation program.
Generally there are two priority levels: normal and emergency calls.
Each call is one of the two types with 1/2 probability. The means
and the standard deviation are presented of 100 samples with 10 000
offers handled in each sample. The further setups for the simulation
can be seen on Table~\ref{table:reqB}.

\begin{table}\label{table:reqB}
\center
\begin{tabular}{|c|c|c||c|c|c|}
\hline
$90$ & $100$ & $W_H=10$ & \{0.,1.\} & \{0.38,0.62\} & \{0.01,0.99\} \\
 &  & $W_L=15$ & [.002,.002] & [.015,.015] & [.006,.006] \\
\hline
$150$ & $100$ & $W_H=10$ & \{0.2,0.98\} & \{0.4,0.6\} & \{0.05,0.95\}\\
 &  & $W_L=15$ &  [.003,.003] & [.007,.007] & [.005,.005] \\
\hline
$10$ & $10$ & $W_H=10$ & \{0.,1.\} & \{0.31,0.69\} & \{0.,1.\}\\
 &  & $W_L=20$ & [.000,.000] & [.014,.014] & [.000,.000] \\
\hline
$10$ & $10$ & $W_H=10$ & \{0.5,0.5\} & \{0.5,0.5\} & \{0.5,0.5\}\\
 &  & $W_L=10$ & [.008,.008] & [.009,.009] & [.007,.007] \\
\hline
\end{tabular}
\caption{In each row the following quantities are presented
respectively: total offer rate: $\rho$, maximal throughput: $c$,
watermark settings: $W_{\mbox{high}},W_{\mbox{low}}$ while
$T_j:=W_j/c$. Then portion in rejected messages for Token Bucket,
rate based call gapping and the mixed mechanism respectively.}
\end{table}

It can be seen that all three methods reject less offer from those
of higher priority but Token Bucket ($\gamma_t$) and the mixed
mechanisms ($\gamma_x$) enforce a more strict priority handling than
the simple proposal. Note that in case of sustained overload (row 2)
almost all dropped offers are the lower priority ones.

\subsection{\reqC}
The results tell explicitly that unlike the new rate based call
gapping proposal the original Token Bucket algorithm does not meet
\reqC. We consider a scenario when there are two traffic classes
Class A and Class B. The agreed share for Class A is the 20\% of the
total capacity of the node while the share for Class B is the
remaining 80\%. The offer rates set for the simulator are exactly
the inverse of this for the two type of traffic.

The aggregate offer rate increases from 0.7 offers/sec to 2
offers/sec and reaches the scenario of 100\% overload (the capacity
of the node is mean 1 offer/sec while the offered rate is a mean 2
offers/sec). The offer rate of traffic Class B is $0.4$ i.e. it is
still under its provided share thus all such calls are admitted. On
the other hand the whole remaining capacity should be granted to
traffic Class A and it should be admitted on a higher level than the
agreed share and only those exceeding the capacity limit are to be
rejected.

\begin{figure}[h]
\begin{center}
\includegraphics[width=0.45\textwidth]{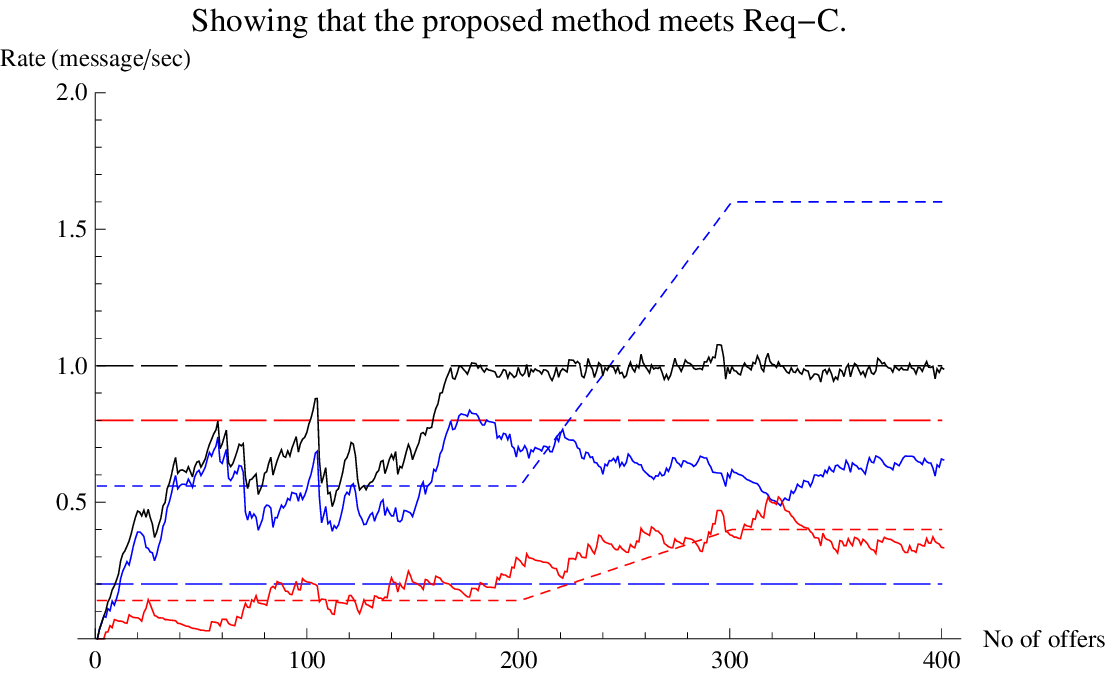}
\end{center}
\caption{The new algorithm ($\gamma_g$) with two traffic
classes.}\label{fig:rbcg-meets-reqc}
\end{figure}

\begin{figure}[h]
\begin{center}
\includegraphics[width=0.45\textwidth]{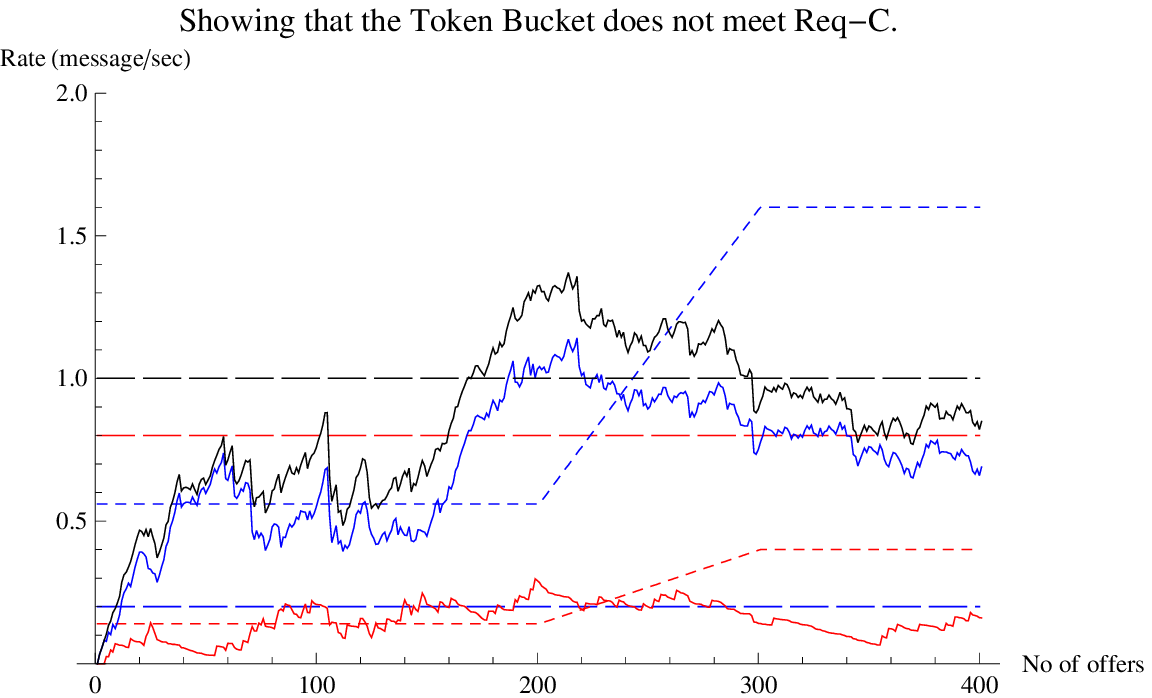}
\end{center}
\caption{The special bucket algorithm that is not able to do meet
any criteria because the bucket size has nothing to do with the
offer and admission rates.}\label{fig:tb-not-meets-reqc}
\end{figure}

Figure~\ref{fig:rbcg-meets-reqc} shows how the behavior of the rate
based call gapping mechanism while one can see the Token Bucket with
exactly the same offered traffic on
Figure~\ref{fig:tb-not-meets-reqc}. (In the simulation we also
implemented a variant of Token Bucket that uses the $\tl$ estimate
and works like the Rate Based Call Gapping as mentioned
in~\ref{section:token-bucket-rate-model-class-handling} but since
the $\tl$ estimate has nothing to do with the intensity of the
traffic the result was the worst of all.)

With the proposed mechanism the minimum share is guaranteed for
traffic Class B (the admission line is around the offered) while the
requirement fails for Token Bucket. With the proposed method there
is no rejected message of Class B since it never offers on a higher
rate than the agreed share. The throughput of the throttle is
limited but also maximized since Class A is granted all remaining
capacity.

\comment{
\subsection{Smoothness}
In this section we show how the different settings of parameter $T$
(parameters $T_j$) affects the smoothness of the figures. As it was
already discussed the rate estimators are asymptotically unbiased as
$T\toinf$ meaning that large $T$ values makes the rate estimation
better and robust against variance. On the other hand small $T$
values have the benefit of reacting more intensively for traffic
rate changes.}

\section{Conclusions}
We have presented the ``rate based call gapping'' mechanism and its
extension with the original Token Bucket mechanism. These unique
mechanisms meet the \textit{maximal throughput with bound}
requirement, handle priorities and give minimum share for different
traffic classes without using message buffers or queues.

Examining the properties of the mechanisms we gave mathematical
definitions of the three requirements and accompanied the
mathematical model with several theorems. Still the proof of
priority handling is missing for the new methods, rather we have
statistical analysis with the simulation we have coded to underpin
our proposal and findings.

Our rate based call gapping strategy can use different traffic
intensity estimators. It is still an open question to find the
optimal estimator or the optimal parameter setting of the estimators
considering Poisson input traffic with variable intensity or even
non-Poisson (e.g. general renewal or Hawkes type) input process.

\section{Appendix}
Notations:\\
\\
\begin{tabular}{|l|l|}
  \hline
  $a,a(t)$ & Real admission rate\\
  $\hat{a},\haa(t)$ & Estimated admission rate\\
  $b,b(t)$ & Actual bucket fill\\
  $c(t)$ & Maximal capacity of the target (rate)\\
  $g_i,g_i(t)$ & Goal rate for traffic class $i$\\
  $g,g(t)$ & Sum of goal rates of all traffic classes\\
  $r, r(t)$ & Token Bucket token generation rate\\
  $T$ & Parameter of the estimator\\
  $T_j$ & Parameter of the estimator for priority level $j$\\
  $u,u(t)$ & Used capacity according to \reqB\\
  $W$ & Watermark for Token Bucket\\
  $W_j$ & Watermark for offers of priority level $j$\\
  $\hat{\alpha},\ha(t)$ & Estimated preliminary admission rate\\
  $\beta,\beta(t)$ & Preliminary bucket size \\
  $\gamma_t$ & The token bucket throttle function\\
  $\gamma_g$ & The rate based call gapping throttle function\\
  $\gamma_{g'}$ & The variant of the rate based\\
  & call gapping throttle function\\
  $\gamma_x$ & The rate based call gapping throttle function\\
  & with Token Bucket extension\\
  $\lambda,\l(t)$ & Intensity (rate) of a Poisson process\\
  $\hat{\lambda},\hl(t)$ & Estimated rate (intensity)\\
  $\rho,\rho(t)$ & Real offer rate\\
  $\hat{\rho},\hr(t)$ & Estimated offer rate\\
  \hline
\end{tabular}

\section{Acknowledgements}
The research was motivated by Ericsson Research Hungary and the High
Speed Network Laboratory and was partially founded by the Hungarian
Ministry of Culture and Education with reference number NK 63066 and
the National Office for Research and Technology with reference
number TS 49835\comment{Fritz Jozsef fele OTKA szam}. Special thanks
to J\'anos T\'oth (Budapest Univ. of Tech. and Eco., Dept. of Math.
Analysis) for the careful reviews and support.

\end{document}